\documentstyle[12pt]{article}
\begin{document}

\def\subsubsection{\@startsection{subsubsection}{3}{\z@}{-3.25ex plus
 -1ex minus -.2ex}{1.5ex plus .2ex}{\large\sc}}
\renewcommand{\thesection}{\arabic{section}}
\renewcommand{\thesubsection}{\arabic{section}.\arabic{subsection}}
\renewcommand{\thesubsubsection}
{\arabic{section}.\arabic{subsection}.\arabic{subsubsection}}
\pagestyle{plain}
\sloppy
\textwidth 150mm
\textheight 620pt
\topmargin 35pt
\headheight 0pt
\headsep 0pt
\topskip 1pt
\oddsidemargin 0mm
\evensidemargin 10mm
\setlength{\jot}{4mm}
\setlength{\abovedisplayskip}{7mm}
\setlength{\belowdisplayskip}{7mm}
\newcommand{\be}{\begin{equation}}
\newcommand{\bel}[1]{\begin{equation}\label{#1}}
\newcommand{\ee}{\end{equation}}
\newcommand{\bea}{\begin{eqnarray}}
\newcommand{\ba}{\begin{array}}
\newcommand{\eea}{\end{eqnarray}}
\newcommand{\ea}{\end{array}}
\newcommand{\noin}{\noindent}
\newcommand{\ra}{\rightarrow}
\newcommand{\txs}{\textstyle}
\newcommand{\disp}{\displaystyle}
\newcommand{\scs}{\scriptstyle}
\newcommand{\scscs}{\scriptscriptstyle}
\newcommand{\sx}[1]{\sigma^{\, x}_{#1}}
\newcommand{\sy}[1]{\sigma^{\, y}_{#1}}
\newcommand{\sz}[1]{\sigma^{\, z}_{#1}}
\newcommand{\sP}[1]{\sigma^{\, +}_{#1}}
\newcommand{\sM}[1]{\sigma^{\, -}_{#1}}
\newcommand{\spm}[1]{\sigma^{\,\pm}_{#1}}
\newcommand{\Spm}{S^{\,\pm}}
\newcommand{\Sz}{S^{\, z}}
\newcommand{\hspf}{\hspace*{5mm}}
\newcommand{\hspt}{\hspace*{2mm}}
\newcommand{\vspf}{\vspace*{5mm}}
\newcommand{\vspt}{\vspace*{2mm}}
\newcommand{\hsix}{\hspace*{6mm}}
\newcommand{\hfour}{\hspace*{4mm}}
\newcommand{\vsix}{\vspace*{6mm}}
\newcommand{\vfour}{\vspace*{4mm}}
\newcommand{\vtwo}{\vspace*{2mm}}
\newcommand{\htwo}{\hspace*{2mm}}
\newcommand{\honecm}{\hspace*{1cm}}
\newcommand{\vonecm}{\vspace*{1cm}}
\newcommand{\htwocm}{\hspace*{2cm}}
\newcommand{\vtwocm}{\vspace*{2cm}}
\newcommand{\ru}{\rule[-2mm]{0mm}{8mm}}
\newcommand{\tinf}{\rightarrow \infty}
\newcommand{\udl}{\underline}
\newcommand{\ovl}{\overline}
\newcommand{\nwl}{\newline}
\newcommand{\nwp}{\newpage}
\newcommand{\clp}{\clearpage}
\newcommand{\simleq}{\raisebox{-1.0mm}
{\mbox{$\stackrel{\textstyle <}{\sim}$}}}
\newcommand{\simgeq}{\raisebox{-1.0mm}
{\mbox{$\stackrel{\textstyle >}{\sim}$}}}
\newcommand{\half}{\mbox{\small$\frac{1}{2}$}}
\newcommand{\smfrac}[2]{\mbox{\small$\frac{#1}{#2}$}}
\newcommand{\bra}[1]{\mbox{$\langle \, {#1}\, |$}}
\newcommand{\ket}[1]{\mbox{$| \, {#1}\, \rangle$}}
\newcommand{\exval}[1]{\mbox{$\langle \, {#1}\, \rangle$}}
\newcommand{\BIN}[2]
{\renewcommand{\arraystretch}{0.8}
\mbox{$\left(\ba{@{}c@{}}{\scs #1}\\{\scs #2}\ea\right)$}
\renewcommand{\arraystretch}{1}}
\newcommand{\UQSU}{\mbox{U$_{q}$[sl(2)]}}
\newcommand{\UQSUA}{\mbox{U$_{q}$[$\widehat{\mbox{sl(2)}}$]}}
\newcommand{\comm}[2]{[\, #1 , \, #2 \, ]}
\newcommand{\intpa}[1]{\mbox{\small $[ \, #1 \, ]$}}
\newcommand{\BC}{boundary conditions}
\newcommand{\TBC}{toroidal boundary conditions}
\newcommand{\ABC}{antiperiodic boundary conditions}
\newcommand{\PBC}{periodic boundary conditions}
\newcommand{\CBC}{cyclic boundary conditions}
\newcommand{\OBC}{open boundary conditions}
\newcommand{\XXZ}{XXZ Heisenberg chain}
\newcommand{\ZFQC}{Zamolodchikov Fateev quantum chain}
\newcommand{\FSS}{finite-size scaling}
\newcommand{\FSSS}{finite-size scaling spectra}
\newcommand{\DT}{duality transformation}
\newcommand{\POM}{Potts models}
\newcommand{\PM}{projection mechanism}
\newcommand{\PQC}{Potts quantum chain}
\newcommand{\TLA}{Temperley-Lieb algebra}
\newcommand{\PTLA}{periodic Temperley-Lieb algebra}
\newcommand{\HA}{Hecke algebra}
\newcommand{\VA}{Virasoro algebra}
\newcommand{\CFT}{conformal field theory}
\newcommand{\SS}{steady state}
\newcommand{\FS}{Fock space}
\newcommand{\FSF}{Fock space formalism}
\newcommand{\ME}{master equation}
\newcommand{\DEX}{diffusion with particle exclusion}
\newcommand{\EVOP}{evolution operator}
\newcommand{\Gtwo}{two-point correlation function}
\newcommand{\au}{\mbox{\"a}}
\newcommand{\uu}{\mbox{\"u}}
\newcommand{\ou}{\mbox{\"o}}
\newcommand{\Au}{\mbox{\"A}}
\newcommand{\Uu}{\mbox{\"U}}
\newcommand{\Ou}{\mbox{\"O}}
\newcommand{\su}{$\beta \hspace*{1.5mm}$}

\begin{titlepage}
\thispagestyle{empty}
\hfill WIS-93/64/Jul.-PH
\begin{center}
\vspace*{1cm}
{\bf \Large
On \mbox{U$_{q}$[SU(2)]}-symmetric Driven Diffusion
}\\[30mm]

{\Large {\sc
Sven Sandow\footnotemark[1] \vtwo \\
and \vfour\\ Gunter Sch\"utz}\footnotemark[2]
} \\[8mm]

\begin{minipage}[t]{13cm}
\begin{center}
{\small\sl
  \footnotemark[1]
  Department of Electronics,
                Weizmann Institute, \\
  Rehovot 76100,
  Israel
  \\[2mm]
  \footnotemark[2]
  Department of Physics,
                Weizmann Institute, \\
  Rehovot 76100,
  Israel}
\end{center}
\end{minipage}
\vspace{20mm}
\end{center}
{\small
We study analytically a model where particles with a hard-core repulsion
diffuse on a finite one-dimensional lattice with space-dependent,
asymmetric hopping rates. The system dynamics are given by the
\mbox{U$_{q}$[SU(2)]}-symmetric
Hamiltonian of a generalized anisotropic Heisenberg antiferromagnet.
Exploiting this symmetry we derive exact expressions for various
correlation functions.
We discuss the density profile and the two-point function and compute
the correlation length $\xi_s$ as well as the correlation time $\xi_t$.
The dynamics of the density and the correlations are shown to be
governed by the energy gaps of a one-particle system.
For large systems $\xi_s$ and $\xi_t$ depend only on the asymmetry.
For small asymmetry one finds $\xi_t \sim \xi_s^2$ indicating
a dynamical exponent $z=2$ as for symmetric diffusion.
}
\\
\vspace{5mm}\\
\udl{PACS numbers:} 05.40+j, 05.60.+w, 75.10.Jm
\end{titlepage}
\newpage
\baselineskip 0.3in

Recently there has been much interest in
stochastic processes the time evolution of which can be mapped to
Hamiltonians of quantum lattice models. A particularly interesting
class of systems (comprising many interesting equilibrium and
non-equilibrium models) are those where the dynamics are mapped to
integrable quantum chains in one dimension \cite{KDN} - \cite{adhr}
 or to
$SU(2)$-symmetric quantum Hamiltonians in any dimension \cite{scsa}.
In these cases the symmetries of the models can be used to determine
critical exponents, correlation functions and other quantities
of interest. Among the
models which can be studied in this way are diffusive
lattice gases (which are related to growth models),
diffusion-limited chemical reactions and vertex models.

In this letter we study a driven lattice gas of particles on
a one-dimensional chain of length $L$. We assume a
hard core repulsion preventing the occupation of each site
by more than one particle.
Particles hop stochastically to their right and left nearest
neighbours with space-dependent hopping rates.
We consider reflective
boundary conditions, i.e.,
particles cannot enter or leave the system. We will show that
if the driving force (represented by an asymmetry in left- and
right-hopping probabilities) is spatially constant the system
has a \mbox{U$_{q}$[SU(2)]} symmetry even if the particle
 mobility has
an arbitrary space dependence. We will use this symmetry
to derive exact results for the correlation functions
even for finite systems.
One should keep in mind that even though we are studying a specific
model, our method and some of the results can be generalized to
other systems. This aspect will be briefly discussed at the end
of this letter.
\vfour \\

In what follows we will study one-dimensional
systems described by occupation numbers
$\udl {n}= \{n_j\}$ where
$j$ is the number of a lattice site und runs from $1$ to $L$. Their
dynamics
are given by a master equation for the probability distribution
$F(\udl{n},t)$ which can
be mapped onto a Schr\mbox{\"o}dinger equation \cite{doi} -
\cite{sa1}:
\bel{A}
\partial_t \ket{F(\udl{n},t)}=H\ket{F(\udl{n},t)}
\htwo .
\ee
We will focus on a system of particles with hard core repulsion
diffusing
on a lattice with position dependent, asymmetric hopping rates.
It shall be subjected to reflecting (sometimes called free) boundary
conditions. The Hamiltonian is then given by:
\bea \label{B}
H&=&\sum_{j=1}^{L-1}[\alpha_j c^{\dagger}_{j}c_{j+1}+\beta_j
c^{\dagger}_{j+1}c_{j}
-\alpha_j c_{j}c^{\dagger}_{j}c^{\dagger}_{j+1}c_{j+1}
-\beta_jc_{j+1}c^{\dagger}_{j+1}c^{\dagger}_{j}c_{j}]\\
\label{B2}
&=& \sum_{j=1}^{L-1} \sqrt{\alpha_j \beta_j}\; u_j
\htwo .
\eea
Here $c^{\dagger}_{j}$ creates a particle at site $j$ and $c_{j}$
annihilates a
particle at site $j$. They obey Pauli commutation relations.
$\alpha_j$ is the
probability rate for a single particle jump from site $j+1$ to site $j$
and $\beta_j$ is the
rate for a jump from $j$ to $j+1$.
The logarithm $f_j=\ln \frac{\alpha_j}{\beta_j}$ of the ratio of the
hopping rates corresponds to a "driving force" while the product
${\alpha_j}{\beta_j}$ is a measure for the mobility of the particles.
Here we assume the force to be independent of the position, i.e.,
\bel{C}
\mbox{e}^{f_j}=\frac{\alpha_j}{\beta_j}=q^2 \htwo \forall \ j \htwo
\mbox{with} \ q
\  \mbox{real}
\htwo .
\ee
$q=1$ corresponds to diffusion without a driving force (symmetric
diffusion).

We would like to point out that the model is related to an integrable
quantum chain. Introducing operators
$ \{\tilde{\sigma}^x_{j},\tilde{\sigma}^y_{j},\tilde{\sigma}^z_{j}\}
=\{q^{-j} c^{\dagger}_{j}+q^{j} c_{j},-i(q^{-j}
c^{\dagger}_{j}-q^{j}c_{j}),1-2c^{\dagger}_{j}c_{j} \} $, which commute
like Pauli-matrices, the Hamiltonian reads: \bel{D}
H=\frac{1}{2} \sum_{j=1}^{L-1}\sqrt{\alpha_j \beta_j}
[\tilde{\sigma}^x_{j} \tilde{\sigma}^x_{j+1}+\tilde{\sigma}^y_{j}
\tilde{\sigma}^y_{j+1} +\frac{q+q^{-1}}{2}\tilde{\sigma}^z_{j}
\tilde{\sigma}^z_{j+1}
+\frac{q-q^{-1}}{2}(\tilde{\sigma}^z_{j+1}-\tilde{\sigma}^z_{j})-
\frac{q+q^{-1}}{2}] \htwo .
\ee
This is the Hamiltonian of a generalized
Heisenberg antiferromagnet. It is the sum of generators $u_j$  of a
Temperley-Lieb
algebra \cite{tla} and it is symmetric under the action of the quantum
group
\mbox{U$_{q}$[SU(2)]} \cite{kr,pasa}.

The generators of
the quantum group
\mbox{U$_{q}$[SU(2)]}
are given in terms of creation and annihilation
operators as:
\be \label{E1}
S^{+}=\sum_{j=1}^{L}b_{j}(q)\;\;,
\;\;S^{-}=\sum_{j=1}^{L}b^{\dagger}_{j}(q)\
\;,\;\;S^{z}=\sum_{j=1}^{L}(-n_{j}+\frac{1}{2})
\ee
with
\bea \label{E2}
b^{\dagger}_j (q)&=&q^{\sum_{k=1}^{j-1}(n_k-1)}
c^{\dagger}_j q^{-\sum_{k=j+1}^{L}(n_k-1)}\;\;,\\
 \label{E3}
b_j (q)&=&q^{\sum_{k=1}^{j-1}n_k}
c_j q^{-\sum_{k=j+1}^{L}n_k}
\eea
and $n_j=c^{\dagger}_jc_j$.
They satisfy the following relations:
\bel{E4}
[S^+,S^{-}]_{-}=[2S^z]_q \;\;\;\mbox{and}\;\;\;
[S^z,S^{\pm}]_-=\pm S^{\pm}
\ee
where $[x]_q=\frac{q^x-q^{-x}}{q-q^{-1}}$ \ .

It is easy to check that each term $u_j$ in (\ref{B2}) commutes with
$S^{\pm}$ and $S^z$. Hence
\begin{eqnarray}
\label{G1}
[H,S^{\pm}]_{-}&=&0\\
\label{G2}
[H,S^{z}]_{-}&=&0
\htwo .
\eea
Since $S^z=L/2 -\sum_{j=1}^L n_j$
eq. (\ref{G2}) states that the particle number is conserved,
which is obvious for a purely diffusive system.
Eq. (\ref{G1}), however, has the nontrivial consequence that we can
construct all the zero energy eigenstates out of the zero particle
state $| \; 0 \; \rangle$. One obtains
\bea
\label{H1}
H| \; N \;\rangle&=&0 \;\;\; \mbox{with} \;\;\;
| \; N \;\rangle=\frac{1}{[N]_q!}(S^{-})^{N} |\; 0 \; \rangle\\
\label{H2}
 \mbox{and}\;\;\; \langle \; N \;|H&=&0 \;\;\; \mbox{with} \;\;\;
 \langle \; N \;|=\frac{1}{[N]_q!}\langle \; 0 \; |(S^{+})^{N}
  \htwo .
 \end{eqnarray}\\
Here $[m]_q!=[1]_q [2]_q  ...
[m]_q$. The left eigenstates
assign equal weight to any $N$-particle configuration.
Hence, averaging over an $N$-particle state is performed by multiplying
$\langle \; N \;|$ from the left.
The normalized $N$-particle states
\begin{equation}\label{I}
| \; N \;\rangle_{norm}=\BIN{L}{N}^{-1}_q
| \; N \;\rangle=\BIN{L}{N}^{-1}_q q^{N(L+1)}
\sum_{\udl{n}\}}^{\;\;\;
\;\;\;\;(N)} q^{-2 \sum_{j=1}^L j n_j}| \; \underline{n} \;\rangle
\end{equation}
are the steady states of the system and satisfy $\langle \; N \; |
\; N \; \rangle = 1$.
Here $\BIN{L}{N}_q=\frac{[L]_q!}{[L-N]_q![N]_q!}$ , and the upper index at
the sum means that the
summation runs only over states with total particle number $N$.

Next we will discuss the properties of the steady state in more
detail.
A configuration with the $N$-particles located at sites $\{
k_1,k_2,...k_N\}$
has a probability proportional to $q^{-2 \sum_{l=1}^N k_l}$ if the system
is in the steady state (\ref{I}). Note that these states obey a datailed
balance condition. Two special cases are easily understood. For $q=1$, i.e.,
for symmetric diffusion, every configuration has the same probability in
accordance with what is known. In the limit $q\rightarrow 0$ we find
 a configuration where the particles occupy the sites
$\{L-N+1,L-N+2,...,L\}$ with probability 1, which is plausible. The case
$q\rightarrow \infty$ is
equivalent to $q\rightarrow 0$ since simultaneously replacing $q$ by
$q^{-1}$ and each position $x$ by $L+1-x$  leaves the system invariant.

For the calculation of
averaged quantities in a state (\ref{I}) the following relations are useful:
\bea
\label{J1}
c_{x}| \; N \;\rangle&=&q^{L+1-2x}(1-n_{x})| \; N-1 \;\rangle\\
\label{J2}
\langle \; N \;|c^{\dagger}_x&=& \langle \; N \;|(1-n_x)
\htwo .
\eea
{}From this we find a recursion for one-time correlation functions:
\bel{K}
\langle \; N \;|n_{x_1}...n_{x_l}| \; N \;\rangle_{norm} =
\frac{[N]_q q^{L+1-2x_l}}{[L-N+1]_q}
\langle \; N-1 \;|n_{x_1}...n_{x_{l-1}}(1-n_{x_l})| \; N-1 \;
\rangle_{norm} \htwo .
\ee

Of particular interest is the density profile   $\rho_N(x)=\langle \; N
\;|n_{x}| \; N \;\rangle_{norm}$  in the $N$-particle
steady state. Solving the recursion (\ref{K}) and inserting
$\rho_1(x)= q^{L+1-2x}/[L]_q$
we find the exact expression:
\bel{L}
\rho_N(x)=\BIN{L}{N}^{-1}_q \sum_{k=0}^{N-1}(-1)^{N-k+1}
q^{(N-k)(L+1-2x)} \BIN{L}{k}_q
\htwo .
\ee

The recursion relation (\ref{K}) simplifies for large $N$ and $L$. The
following approximations are then derived (We assume $q<1$.):
\bea \label{M1}
\rho_N(L-N+1-r)&=&q^{2r}+o(q^{4r}) \approx e^{-r/ \xi_s} \\
\label{M2}
\rho_N(L-N+r)&=&1-q^{2r}+o(q^{4r}) \approx 1-e^{-r/ \xi_s}
\;\;\;\mbox{for}\; r=1,2,...\;\;.
\eea
with the decay length
\bel{O1}
\xi_s=\frac{1}{ \ln q^{-2}}
\htwo .
\ee
{}From (\ref{M1}) and (\ref{M2}) the following symmetry becomes obvious:
\bel{N}
\rho_N(L-N+1-r) \approx 1-\rho_N(L-N+r)
\htwo .
\ee
In this approximation the density profile is invariant with respect
to the scale transformation $N\rightarrow bN\;,\;L\rightarrow bL\;,\;
r\rightarrow br\;,\;\xi_s\rightarrow b \xi_s$.
Rescaling only $N,L,r$ and keeping $q$ fixed we obtain a theta-function for
the density profile in the limit $b\ra \infty$.
I.e., a given system observed on a large scale gives
$\rho(x) \approx
\theta (x-L+N)$. Using a higher resolution we find this function to
be smeared. The decay length $\xi_s$ is a measure for the width of the
transition zone seperating the high density region with $\rho=1$ from the
zero density domain. Its center (where $\rho \approx 1/2$)
is located between sites $L-N$ and $L-N+1$.
Note that $\xi_s$  does not depend on $N$ or $L$.

Another important quantity is the connected
two-point correlation which is defined as
\bel{N1}
C(x,y,t)=\langle \;n_x(t)n_y(0)\;\rangle_N -
\rho_N(x)\rho_N(y)
\htwo .
\ee
where averaging is performed over the $N$-particle steady state.

For the one-time correlation function we find an exact expression
solving the recursion relation (\ref{K}):
\bel{O}
C(x,y,0)=
\frac{q^{2y}\rho_N(y)-q^{2x}\rho_N(x)}{q^{2y}-q^{2x}}
-\rho_N(x)\rho_N(y)
\htwo .
\ee
It is interesting to note that as in a similar diffusion model
with open boundary conditions, where particles are injected and
removed \cite{2} - \cite{DE},
the two-point function can be expressed in
terms of one-point functions.
The two-point function for a large system has a finite amplitude
 only in the transition zone of the
density profile. Inserting the expressions (\ref{M1}) and (\ref{M2})
it turns out that the correlation length
is identical to the decay
length $\xi_s$ (see eq. (\ref{O1})) of the density.
\vfour\\

In order to understand the dynamics of the system we study the
time-dependent correlation function (\ref{N1}).
Using the results for $\rho_N(x)$ it is easy to see  that
like the one-time correlator it is
nonzero only in the transition zone of the density profile. A more
detailed analysis starts with the explicit form following from
(\ref{N1}):
\bel{P}
C(x,y,t)= \BIN{L}{N}^{-1}_q
\langle \;N\;|\;n_x\,\mbox{e}^{Ht}\,n_y\;|\;N\;\rangle
- \BIN{L}{N}^{-2}_q
\langle \;N\;|\;n_x\;|\;N\;\rangle \langle \;N\;|\;n_y\;|\;N\;\rangle
\htwo .
\ee
Inserting a complete set of energy eigenstates, defined by
$H\;|\;k\;\rangle=\epsilon_k\;|\;k\;\rangle$ and $
\langle \;k\;|\;H=\langle \;k\;|\; \epsilon_k$  we get:
\bea \label{Q1}
C(x,y,t)= \BIN{L}{N}^{-1}_q
\sum ^{\;\;\;\;\;\;\; \prime}_k \frac{A_k(x)B_k(y)}
{\langle \;k\;|\;k\;\rangle} \;
\mbox{e}^{\epsilon_k t} \\
\label{Q2}
\mbox{with}\;\;\;
A_k(x)=\langle \;N\;|\;n_x\;|\;k\;\rangle\;\;\;,\;\;\;
B_k(y)=
 \langle \;k\;|\;n_y\;|\;N\;\rangle
\eea
where the prime means that the steady state is excluded from the
summation. (Its contribution cancels with  the second term in
(\ref{P}).)
Because of the
\mbox{U$_{q}$[SU(2)]}
symmetry the energy eigenstates are eigenstates of
$S^z$ and the Casimir operator
$S^2=2S^+S^-+[S^z]_q[S^z+1]_q$.
 Hence, the energy
eigenstates $|\;k\;\rangle$ can be classified by their spin-componenet $S^z$
and by their total spin $S$.
Since for real $q$ the representations of
\mbox{U$_{q}$[SU(2)]}
are isomorphic to the
ones of $SU(2)$ \cite{kr,pasa}
the selection rules for matrix elements are the
same for both symmetries. Consequently, the only nonvanishing matrix
elements (\ref{Q2}) are those with states $|\;k\;\rangle$ which have
$S=L/2-1$ and $S^z=L/2-N$ \cite{stin}. These states are given by
$| \;N,k \;\rangle=\frac{1}{[N]_q!}(S^{-})^{N-1} |\;1,k\; \rangle$ ,
$ \langle \;N,k\;|=\frac{1}{[N]_q!}\;   \langle \;1,k\; |(S^{+})^{N-1}$,
respectively, where we have introduced one-particle eigenstates  as:
\bel{R}
H\;|\;1,k\;\rangle
=\epsilon_k\;|\;1,k\;\rangle    \;\;\; \mbox{and}\;\;\;
  \langle \;1,k\;|\;H=\;  \langle \;1,k\;|\; \epsilon_k
\htwo .
\ee
Hence, the time dependence of the correlation
functions is determined by the
eigenenergies of the one-particle system, which are much easier to
calculate than many-particle energies. The matrix elements (\ref{Q2})
reduce to the one-particle matrix elements
$A_k(x) = \langle \, 1 \, | (S^{+})^{N-1} \, n_x \,
(S^{-})^{N-1} \, | \, 1,k \, \rangle / ([N]_q!)^2$ and
$B_k(y) = \langle \, 1,k \, | (S^{+})^{N-1} \, n_y \,
(S^{-})^{N-1} \, | \, 1 \, \rangle / ([N]_q!)^2$.
The exact form of the correlation
function can be obtained by evaluating them which
(if the eigenstates $| \, 1,k \, \rangle$ are known) is a
tedious but straightforward task.

The time evolution of the density profile can be studied in
the same way as the correlation functions.
If the system starts with an arbitrary initial $N$-particle state
$|\;F(0)\;\rangle$ the density is
$\rho(x,t)=\langle\;N\;|\; n_x \;
\mbox{e}^{Ht}\;|\;F(0)\;\rangle$. It decays to its
stationary value (\ref{L})
on time scales given by the one-particle energies
(\ref{R}). Expanding $|\;F(0)\;\rangle$ in eigenstates of $H$
one finds that contributions to the time dependence of $\rho(x,t)$
come from those
excitations $\epsilon_k$ for which the matrix
elements $     \langle\;1,k\;|\; (S^+)^{N-1}\;|\;F(0)\;\rangle$
do not vanish.

Up to this point our discussion was completely general and applies
actually to any system of exclusive particles with an
\mbox{U$_{q}$[SU(2)]}-symmetric time evolution operator.
In what follow we will discuss the system (\ref{B}) with homogeneous
jump rates $\alpha_j=q\;,\;\beta_j=q^{-1}\;\;\;
\forall j$. For this case we
find the eigenenergies:
\bel{U}
\epsilon_k=2 \cos (\frac{2 \pi}{L}k)-(q+q^{-1})<0
\;\;\;\mbox{for}\;\;\;k=1,2,...L-1
\htwo .
\ee
The long-time behaviour of the correlation function (\ref{Q1})
(and of the density $\rho(x,t)$ if
the initial state is not orthogonal to the state
$|\;N,k_1\;\rangle$)
is governed by the lowest
excitation $\epsilon_1$. Consequently, we can identify
\bel{W}
\xi_t=-\epsilon^{-1}_1 =
\{ q+q^{-1}  - 2 \cos (\frac{2 \pi}{L}) \}^{-1}
\stackrel{L \ra \infty}{ \longrightarrow } \{ q+q^{-1} -2 \}^{-1}
\ee
with the correlation time of the system.
Note that $\xi_t$ does not depend on $L$ or $N$ if $L\gg 1$.
We would like to emphasize that for ($q\ne1$) $\xi_t$ remains
finite for $L\rightarrow \infty$, as opposed
to the correlation time
for asymmetric diffusion with {\em periodic} boundary
conditions which diverges as $L^{3/2}$
\cite{GS}. The significance of this remark lies in the fact
that it demonstrates the impact of
a boundary term which
alone is sufficient to change the gap structure
of the Hamiltonian. Also for symmetric diffusion $\xi_t$ diverges,
but as $L^{2}$, independent of the boundary term.
If the asymmetry is small, i.e., if $q-1 \ll 1$, one derives from
the expressions (\ref{O1}) and (\ref{W}) that correlation time
and correlation length are related as
$\xi_t \sim \xi^{2}_s$ indicating a dynamical exponent $z=2$.
\vfour\\

Finally, we briefly discuss possible generalizations of our results
and methods to other
\mbox{U$_{q}$[SU(2)]}-symmetric models. Instead of the Hamiltonian
(\ref{B}) one could, for example, study an asymmetric 6-vertex model
with space-dependent vertex weights defined by
the diagonal-to-diagonal transfer matrix
$T=(\prod_{j=1}^{L/2-1} T_{2j})(\prod_{j=1}^{L/2}T_{2j-1})$ with
$T_j = 1 - \sqrt{\alpha_j\beta_j} \, u_j$ \cite{KDN,scsa,sch}.
Here the hopping probabilities $\alpha_j$ and $\beta_j$ parametrize
the vertex weights and $T$ defines a discrete time evolution
with a parallel updating mechanism. In the mapping to the vertex
model the density correlations discussed here correspond to
arrow correlations in the plane. Other models of interest
for which our methods could successfully be
applied are the $q$-deformed
spin-1 Heisenberg chain \cite{bmnr}
describing reaction-diffusion processes
involving two different kinds of exclusive particles or some of the
models discussed in \cite{adhr}. Of course it also possible to
generalize our procedure to models with other $q$-deformed symmetries.
A systematic approach to the study of correlation functions of
stochastic systems
using symmetries of the type discussed here will be presented
elsewhere.
\vfour\\

We would like to
thank H. Spohn and E. Domany for stimulating discussions.
Financial support by the Minerva
foundation (S.S.) and the Deutsche Forschungsgemeinschaft (G.S.) is
gratefully acknowledged.

\bibliographystyle{unsrt}

\end{document}